# Translocation time of a polymer chain through an energy gradient nanopore

Meng-Bo Luo[1,‡], Shuang Zhang[1,2], Fan Wu[1], Li-Zhen Sun[3]

[1] *Department of Physics, Zhejiang University, Hangzhou 310027, China*

[2] *College of Science, Qinzhou University, Qinzhou 535011, China*

[3] *Department of Applied Physics, Zhejiang University of Technology, Hangzhou 310023, China*

*Corresponding author. E-mail: [‡]luomengbo@zju.edu.cn*

The translocation time of a polymer chain through an interaction energy gradient nanopore was studied by Monte Carlo simulations and the Fokker-Planck equation with double-absorbing boundary conditions. Both the simulation and calculation revealed three different behaviors for polymer translocation. These behaviors can be explained qualitatively from free-energy landscapes obtained for polymer translocation at different parameters. Results show that the translocation time of a polymer chain through a nanopore can be tuned by suitably designing the interaction energy gradient.

## 1 Introduction

Translocation of a polymer chain through nanopores has attracted much attention in recent years. One reason is that the phenomenon of polymer translocation is ubiquitous and critically important in physics, chemistry, and biology; examples include the transport of proteins through membrane nanopores [1, 2], translocation of RNA across nuclear pores [3], DNA molecule transfer from virus to host cell, and gene transport between bacteria [4]. Another reason is that the translocation of polymers or biomolecules through nanopores has great potential applications in DNA separation [5, 6], gene therapy, drug delivery, gel electrophoresis [7], and so on.

Since the first experiment on the translocation of a DNA chain through a nanopore in 1996 [8], great progress in nanopore-based sensing and sequencing has been made during the last two decades [9-16]. The idea of this technology is that the ionic current across a nanopore is dependent on the structure and sequence of charged DNA/RNA molecules driven through the nanopore under external electrical potential. The different levels of ionic current as well as different translocation times due to different structures or ingredients of DNA/RNAs will help discriminate different polymer chains [17-21].

Stimulated by the experimental progress, researchers have studied the translocation of polymer chains through nanopores extensively in theory [22-27] and simulations [28-40]. Assuming that the translocation time is much longer than the relaxation time of the polymer, polymer translocation can be treated as a quasi-equilibrium process and the free-energy landscape of the translocation can be obtained by different methods [23, 25, 27, 41]. Based on the free-energy landscape, the translocation time can be derived from the Fokker-Planck equation [22, 23]. To simplify, the diffusion coefficient $D$ of the polymer, which is involved in the Fokker-Planck equation, is usually assumed to be a constant during the translocation. Simulations showed that the translocation time was strongly dependent on the polymer-pore interactions as the free-energy landscape was dependent on the polymer-pore interactions [31, 33, 41]. The interaction between a DNA chain and a nanopore could be controlled by varying the surface charges of the nanopore [13]. A recent experiment found that translocation of the DNA chain can be slowed by modulating the surface charges [42].

Experiment found that the translocation time of a DNA chain could be tuned by modulating the pH gradient across an α-hemolysin protein pore [13]. The result was interpreted as a change in the DNA-pore interaction owing to the pH gradient, since the surface charges of the protein pore can be tuned via protonation of the charged amino acid residues of the protein pore. In our recent theoretical calculation, we assumed that the gradient pH conditions might lead to a gradient polymer-pore interaction [43]. Then the gradient polymer-pore interaction was expressed as

$$E = E_0 + kx \quad , \tag{1}$$

where $E_0$ is the initial potential energy at the entrance of the pore, $x$ is the position of the monomer inside the pore, and $k$ is the energy gradient. Based on the free-energy landscape, the mean first passage time was obtained by applying the Fokker-Planck equation with reflecting-absorbing boundary conditions. We found that the first passage time was dependent on the energy gradient and the time would reach a minimum at a proper gradient interaction [43]. Besides the pH value, the interaction between DNA and the nanopore could be changed by surface modification [44, 45] or by surface-biased voltage [46]. It was pointed out that surface modification would be important to identify different nucleotides [47].

It is therefore important to control or tune the translocation time of the polymer using a gradient nanopore. In this work, the translocation time for a polymer translocating through the gradient nanopore was studied using the Fokker-Planck equation with double-absorbing boundary conditions and Monte Carlo (MC) simulations. Both simulation and calculation results show that the translocation time is dependent on the interaction energy gradient. At proper polymer-pore interaction parameters, the translocation time reaches a minimum. Moreover, three different behaviors — dependent on the energy gradient and driving force — are observed for the translocation time. Results show that the translocation time of a polymer chain through a nanopore can be tuned by appropriately designing the interaction energy gradient and driving force.

## 2 Model and method

A sketch of our system is presented in Fig. 1. An impenetrable membrane separates the whole space into a *cis* side and a *trans* side. The polymer is represented by a chain of $N$ identical linked beads of diameter $D = 1$. The polymer is initially placed at the *cis* side. In this study, the radius of the nanopore is set as $R = 1$, and the length of the nanopore is set as $L = 10$. The narrow nanopore only allows single file translocation of the polymer. We consider a charged polymer and an applied electrical field only exists in the nanopore. The electric field $E$ outside the nanopore can be neglected because $E$ is inversely proportional to the area through which the current flows. Thus any bead inside the nanopore experiences a uniform electrical force $f$. The gradient potential given by Eq. (1) is also taken into account. We therefore study the time for the polymer chain to thread through the cylindrical nanopore to enter the *trans* side under the driving force.

In this work, we study the case where $N > L$. Figure 1 presents the starting state and the ending state for the translocation. The whole translocation process of the polymer can be divided into three stages: (i) *filling stage*, (ii) *transferring stage*, and (iii) *escaping stage*. During the filling stage, the head (H) monomer of the polymer moves from the entrance (C) of the nanopore to the exit (D). During the transferring stage, the tail (T) monomer arrives at C. During the escaping stage, the polymer finally leaves the nanopore as T monomer arrives at D.

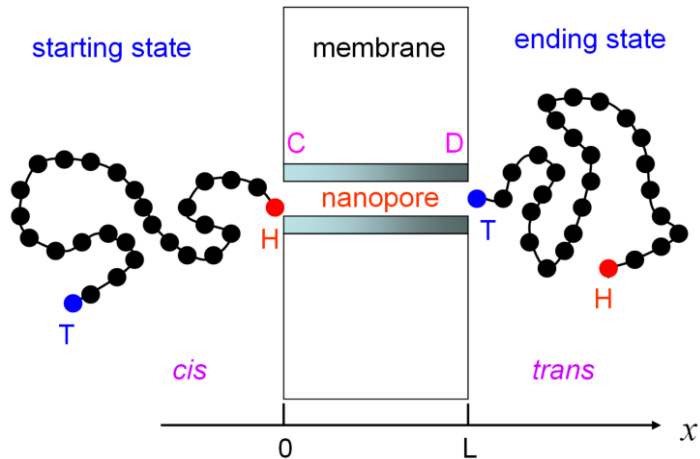

**Fig. 1** A sketch of our simulation model for a polymer translocating through an energy gradient nanopore from the *cis* side to the *trans* side. The starting state and the ending state of the translocation are presented.

The state of the polymer is described by a virtual coordinate $x_v$ for convenience. The virtual coordinate $x_v$ is defined as the position of the head monomer if the whole polymer is fully stretched straight along the nanopore. Therefore, we have $x_v = 0$ for the starting state, and $x_v = N + L$ for the ending state.

The free energy of the polymer during translocation is expressed as a function of the virtual coordinate $x_v$ [43]. The free energy of the polymer is estimated using the formula $F = U - TS$, where $U$ represents the internal energy and $S$ represents the conformation entropy. The internal energy $U$ is composed of two parts: the interaction energy $U_I$ and potential energy $U_P$. The interaction energy $U_I = \sum_i E(x_i)$ is a summation of all polymer-nanopore interactions. The potential energy $U_P = -\sum_i f \cdot \Delta x_i$ is attributed to the driving force $f$, where $\Delta x_i$ is the incremental distance of the $i^{th}$ monomer inside the nanopore. The conformation entropy is given by $S = k_B \ln \Omega$ with $\Omega$ being the conformation number of the polymer. The polymer chains at the *cis* side and at the *trans* side can be treated as end-grafted chains. We use the conformation number $\Omega_n = \mu^n n^{\gamma-1}$ for a chain outside the nanopore and $\Omega = 1$ for that inside the nanopore. To compare with our simulation results, we use $\mu = 20$ for our model polymer chain and $\gamma = 0.69$ for the 3D polymer model [48].

Therefore, the free energy of the polymer can be expressed as

$$F(x_v) = \begin{cases} m(E_0 + kx^*) - mx^* f - \ln \Omega_{N-m} & \text{filling} \\ m(E_0 + kx^*) - (x^* m + sL)f - (\ln \Omega_s + \ln \Omega_{N-s-m}) & \text{transferring}, \\ m(E_0 + kx^*) - [NL - m(L - x^*)]f - \ln \Omega_s & \text{escaping} \end{cases} \quad (2)$$

where $m$ is the number of monomers inside the nanopore, $s$ is the number of translocated monomers at the *trans* side, and $x^*$ is the mean travel distance for these monomers inside the nanopore. We have $s = 0$ for the filling stage and $m + s = N$ for the escaping stage. Further, we have (i) $m = x_v + 1$ and $x^* = \dfrac{x_v}{2}$ for the filling stage, (ii) $m = L + 1$ and $x^* = \dfrac{L}{2}$ for the transferring stage, and (iii) $m = N + L - x_v$ and $x^* = L - \dfrac{1}{2}(N + L - x_v - 1)$ for the escaping stage. The free energy is expressed in units of $k_B T$ with $k_B$ the Boltzmann constant and $T$ the temperature.

## 2.1 Fokker-Planck equation

Utilizing the free-energy landscape $F(x_v)$, the translocation time for the polymer chain moving from the *cis* side to the *trans* side can be calculated. The diffusion of the polymer chain can be described by the Fokker-Planck equation [22]

$$\frac{\partial}{\partial t} p(x_v, t) = L_{FP}(x_v) p(x_v, t) \tag{3}$$

where $p(x_v, t)$ is the probability distribution and the $L_{FP}(x_v)$ is the Fokker-Planck operator described by

$$L_{FP}(x_v) = \frac{1}{l_0^2} \frac{\partial}{\partial x_v} D(x_v) e^{-F(x_v)/k_B T} \frac{\partial}{\partial x_v} e^{F(x_v)/k_B T} . \tag{4}$$

Here $D(x_v)$ is the diffusion coefficient of the whole chain. In this work, we assume that $D(x_v)$ is independent of the coordinate $x_v$ during the translocation, i.e., $D(x_v) = D$.

The translocation time $\tau$ is calculated using the Fokker-Planck equation with absorbing-absorbing boundary conditions [49], i.e., adsorbing boundary conditions at $x_v = 0$ and $x_v = N + L$ are used. The translocation time can be calculated by

$$\tau = \frac{g(0,1) h(1, N+L) - h(0,1) g(1, N+L)}{g(0,1) g(0, N+L)} . \tag{5}$$

Here the step length for polymer moving inside the nanopore is assumed to be $l_0 = 1$. $g(x, y)$ and $h(x, y)$ capture the information regarding the free energy landscape and can be calculated by

$$g(x, y) = \sum_{i=x}^{y} \exp[F(i) - F(0)] / k_B T \tag{6}$$

and

$$h(x, y) = \sum_{i=x}^{y} \sum_{j=0}^{i} \sum_{k=0}^{j} \exp[F(i) - F(j) + F(k) - F(0)] / k_B T . \tag{7}$$

In the calculation using the Fokker-Planck equation, the translocation time $\tau$ is expressed in units of $l_0^2/D$.

## 2.2 Monte Carlo simulation

A coarse-grained off-lattice bead spring polymer model is adopted in the MC

simulation [32]. The interaction between bonded beads is described by the finitely extensible nonlinear elastic potential (FENE):

$$U_{FENE} = \begin{cases} -\dfrac{k_F}{2}(b_{max}-b_{eq})^2 \ln\left[1-\left(\dfrac{b-b_{eq}}{b_{max}-b_{eq}}\right)^2\right] & \text{for } b_{min}<b<b_{max} \\ \infty & \text{otherwise} \end{cases} \quad (8)$$

where $b$ is the bond length, $k_F = 40$ is the elastic constant, $b_{eq} = 0.7$ is the equilibrium bond length, and $b_{max} = 1$ is the maximum bond length. The minimum bond length is $b_{min} = 2b_{eq} - b_{max} = 0.4$. The van der Waals interaction between two non-bonded beads separated by distance $r$ is given by the Morse potential

$$U_M(r) = \varepsilon_M \{\exp[-2\alpha(r-r_{min})] - 2\exp[-\alpha(r-r_{min})]\} \quad (9)$$

with $\alpha = 24$, $r_{min} = 0.8$, and $\varepsilon_M/k_BT = 1$.

The standard Metropolis algorithm scheme is employed to determine the moves of the polymer. In principle, a randomly chosen monomer attempts to move a displacement of $\Delta x$, $\Delta y$, and $\Delta z$ in three directions, where the displacement ($\Delta x$, $\Delta y$, $\Delta z$) is ranged randomly from $-0.5$ to $0.5$ [32, 37]. The trial move is accepted with the probability $p = \min[1, \exp(-\Delta U/k_BT)]$, with $\Delta U$ being the energy shift due to the move. One Monte Carlo step (MCS), which can be scaled to a real time unit, is defined as the time duration in which $N$ monomers are attempted to move once.

We simulated the translocation of a polymer chain from the starting state to the ending start as shown in Fig. 1. Usually, many translocations are attempted before a final successful translocation occurs, where the polymer entirely enters the *trans* side. In other words, the polymer chain can be drawn back even if some monomers have entered the nanopore. Simulation is repeated until the whole chain enters the *trans* side. The trial time $t_{trial}$ is the simulation time cost for all attempted translocations. The duration of the final successful translocation is defined as the translocation time $\tau$. The translocation probability is defined as

$$P = \dfrac{1}{1+n_{trial}} \quad (10)$$

with $n_{trial}$ being the number of attempted translocations before the final successful translocation.

In our simulation, we did not consider ion-ion correlation. Ion-ion correlation is important for the ion current [47], but it would not affect the translocation time appreciably. To save computation time, we therefore did not introduce ions in the system. If ions are introduced in the system, most of the simulation time will be spent on the update of the ion positions since the ions move significantly faster than the polymer chain.

## 3. Results and discussion
### 3.1 Translocation time calculated by Fokker-Planck equation

Figure 2 presents free-energy landscapes for polymer chain translocation through different nanopores under different driving forces. Here, the polymer length $N = 64$ and initial interaction $E_0 = -2$ are fixed. The free energy at $x_v = 0$ is only dependent on $N$, so it is a constant in Fig. 2. The free-energy drop for the translocation $F(x_v = N + L) - F(x_v = 0) = -NLf$ increases with $f$. Therefore the descent slope in the middle region of the free-energy landscape also increases with $f$.

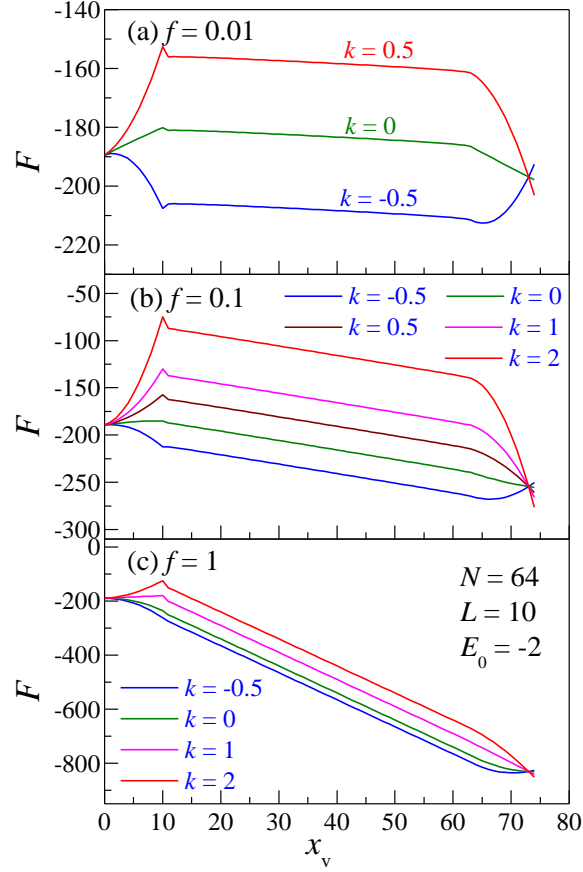

**Fig. 2** Plot of free-energy landscape for polymer translocation through nanopore with different energy gradients at driving force $f = 0.01$ **(a)**, $f = 0.1$ **(b)**, and $f = 1$ **(c)**. Polymer length $N = 64$, nanopore length $L = 10$, and initial attraction $E_0 = -2$.

At weak driving force ($f = 0.01$), the energy drop is small so the free-energy landscape is mainly determined by $k$. The free-energy landscape for translocation is highly dependent on the energy gradient $k$. At $k = -0.5$, the attraction strength increases with translocation, resulting in a wide free-energy well for the translocation. At positive $k$ the attraction strength decreases with increasing $x_v$ and the polymer-pore interaction can be changed to a repulsive interaction at the rightmost position of the nanopore. In this case, a free-energy barrier appears and the barrier increases with $k$.

The free-energy landscape is also highly dependent on the driving force $f$. For the same $k$, the width and depth of the free-energy well decrease with $f$ and the height of the free-energy barrier decreases with increasing $f$. Thus the translocation time decreases with increasing $f$, as expected.

Using the Fokker-Planck equation, we calculated the translocation time $\tau$ for polymers through nanopores with different energy gradients $k$. Similar to the free-energy landscape, $\tau$ is highly dependent on $k$. Fig. 3 presents the dependence of $\tau$ on $k$ for $E_0 = -2$ and $-4$ at $f = 0.01$, $0.1$, and $1$. Generally speaking, $\tau$ decreases with increasing $f$, as expected. The fact that $\tau$ becomes large at negative $k$ is due to the free-energy well for the translocation [43].

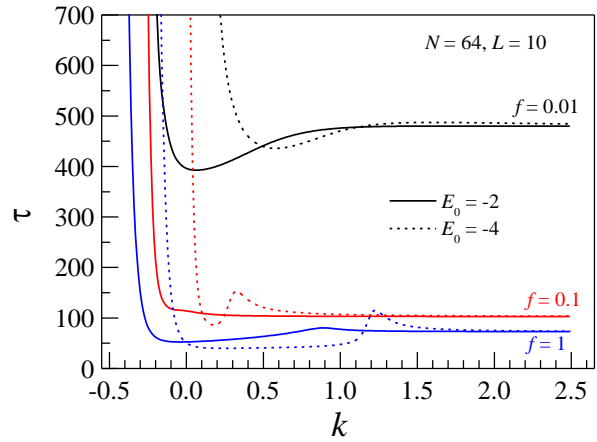

**Fig. 3** Plot of the translocation time $\tau$ for polymer through nanopores of different energy gradient $k$ for three driving forces $f = 0.01$, $0.1$, and $1$. Solid line and dotted line represent the initial interaction $E_0 = -2$ and $-4$, respectively.

It is interesting to find that there are three different behaviors of the translocation time. The first one is that, with increasing $k$, $\tau$ first decreases and then increases, and finally reaches a constant as shown for $f = 0.01$. Figure 2 shows that the free-energy well at negative $k$ changes to a free-energy barrier at positive $k$. At a proper energy gradient, there is neither a significant free-energy barrier nor a free-energy well during the translocation. In this situation, the polymer could translocate through the nanopore fast [41]. At large $k$, we find the free-energy barrier increases with $k$, thereby increasing the translocation time. However, we find the descent slope at the end of translocation increases too, leading to a decrease in the translocation time. Thus, it is reasonable to find a constant translocation time at large $k$.

The second behavior is that $\tau$ decreases monotonously with increasing $k$ as shown for $f = 0.1$ at $E_0 = -2$. The third behavior is that a small peak of $\tau$ appears at large $k$ as shown for $f = 0.1$ and $f = 1$ at $E_0 = -4$. The small peak at large $k$ resulted from the competition between the free-energy barrier and the descent slope of the free-energy at the end of translocation. From Fig. 2(c), it is clear that a large barrier is always accompanied by a sharp downward slope.

**3.2 Monte Carlo simulation results**

Simulations were carried out for a polymer of length $N = 64$ threading a nanopore length of $L = 10$. The mean translocation time $\langle\tau\rangle$ was calculated by performing 1000 independent samples. In addition, the duration times for the filling stage, transferring stage, and escaping stage were calculated separately.

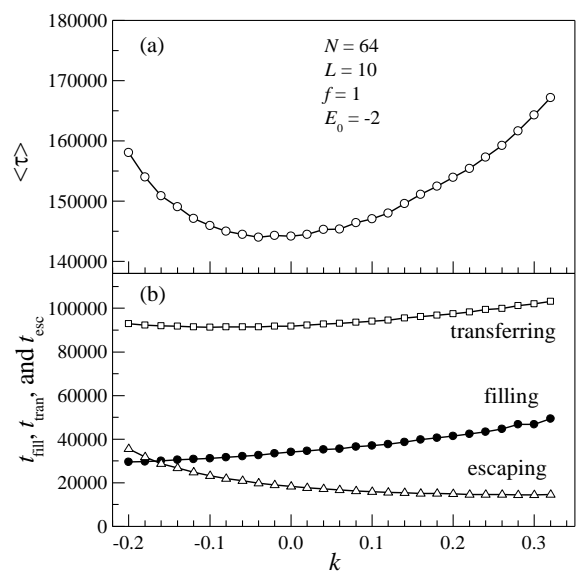

**Fig. 4** Plot of the translocation time $\langle\tau\rangle$ **(a)** and the filling time $t_{\text{fill}}$, transferring time $t_{\text{tran}}$, and escaping time $t_{\text{esc}}$ **(b)** versus the energy gradient $k$ for polymer length $N = 64$. Pore length $L = 10$, driving force $f = 1$, and initial interaction energy $E_0 = -2$.

Figure 4(a) presents the dependence of the mean translocation time $\langle\tau\rangle$ on the energy gradient $k$ when $E_0 = -2$ and $f = 1$. The translocation time showed a minimum value at about $k = -0.04$; such a behavior is similar to that shown in Fig. 3 for $f = 0.01$. The simulation time for $f = 0.01$ is too long, so we only simulated the translocation for $f = 1$.

The time for the transferring stage was roughly the same as shown in Fig. 4(b). However, the filling time and escaping time are dependent on the energy gradient $k$. That is, the polymer-pore interaction significantly affects the filling and escaping of the polymer. The filling time $t_{fill}$ increased with $k$ while the escaping time $t_{esc}$ decreased with increasing $k$, resulting in the minimum of $<\tau>$ at moderate $k$. At negative $k$, the attraction energy increased with the translocation. In this case, it is easy for the polymer to enter the nanopore but difficult to escape from the nanopore. At positive $k$, the attraction energy decreases with the translocation. In this case, it is difficult for the polymer to enter the nanopore but easy to escape from the nanopore. This is in agreement with the free-energy landscape presented in Fig. 2(a).

The dependence of the translocation probability $P$ and the trial time $t_{trial}$ on the energy gradient $k$ are shown in Fig. 5. It is clear that $P$ decreases with increasing $k$ but $t_{trial}$ increases with $k$. The results clearly show that it becomes difficult for the polymer to enter the pore with increase in $k$. That means an attraction between the polymer and nanopore is good for a polymer to enter the nanopore.

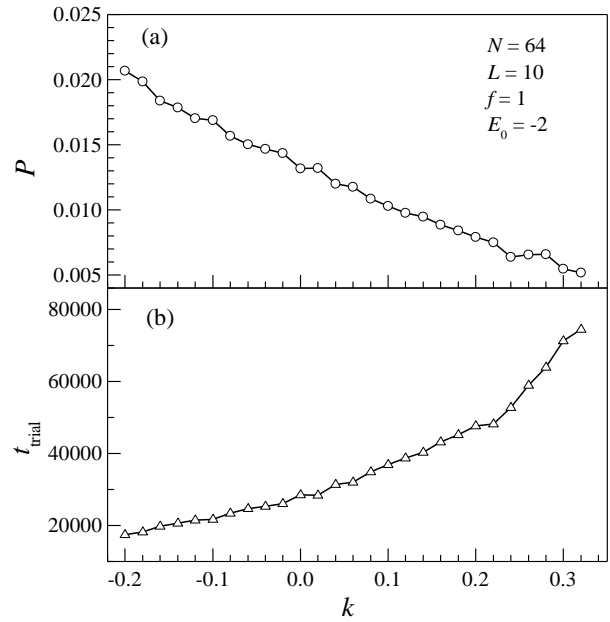

**Fig. 5** Plot of the translocation probability $P$ (**a**) and the trial time $t_{\text{trial}}$ (**b**) versus the energy gradient $k$ for polymer length $N = 64$. Pore length $L = 10$, driving force $f = 1$, and initial interaction energy $E_0 = -2$.

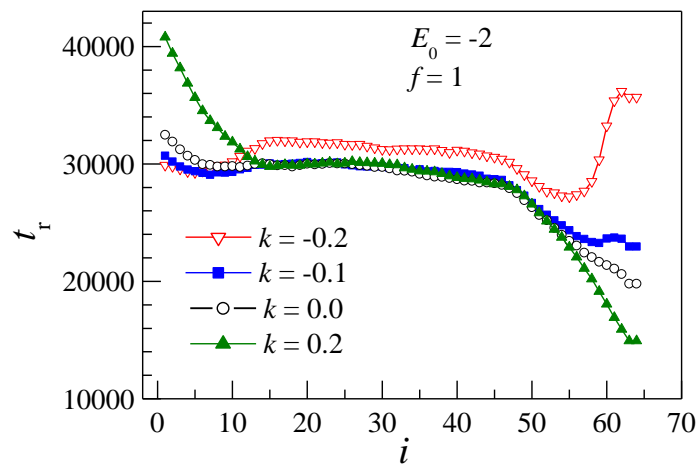

**Fig. 6** Plot of the residence time $t_r$ versus monomer index $i$ for polymer translocation at different energy gradients $k = -0.2, -0.1, 0$, and $0.2$. Polymer length $N = 64$, pore length $L = 10$, driving force $f = 1$, and initial interaction energy $E_0 = -2$.

To investigate the details of the translocation process, we also calculated the residence time $t_r$ for every monomer. The residence time is defined as the duration time of a monomer staying inside the nanopore during the final successful translocation. The residence times at different energy gradients are plotted in Fig. 6. At $k = -0.2$, the attraction energy increased with the translocation. In this case, the residence time is small for the first several monomers but large for the last several monomers. Therefore it is easy for the polymer to enter the nanopore but difficult to escape from the nanopore. With the increase in $k$, the residence times for several head monomers increased whereas those for several tail monomers decreased. At $k = 0.2$, the attraction energy decreased obviously with the translocation. In this case, the residence time was large for the first several monomers but small for the last several monomers. Therefore it is difficult for a polymer to enter the nanopore but easy to escape from the nanopore. The results are consistent with the behavior of the translocation time at different $k$.

As observed from the results of the Fokker-Planck equation, the dependence of $\tau$ on $k$ is dependent on parameters $E_0$ and $f$. There are three different behaviors of $\tau$ as shown in Fig. 3. Our simulation also finds these three behaviors (Fig. 4 and Fig. 7).

In all these three cases, the translocation time $\tau$ is big at small $k$ while the trial translocation time $t_{trial}$ always increases with k. Thus it is very difficult to simulate the translocation of the polymer at small $k$ as well as at large $k$. Our study shows that an interaction gradient nanopore can accelerate or decelerate the translocation of the polymer. It would provide a method to control the translocation time of the polymer.

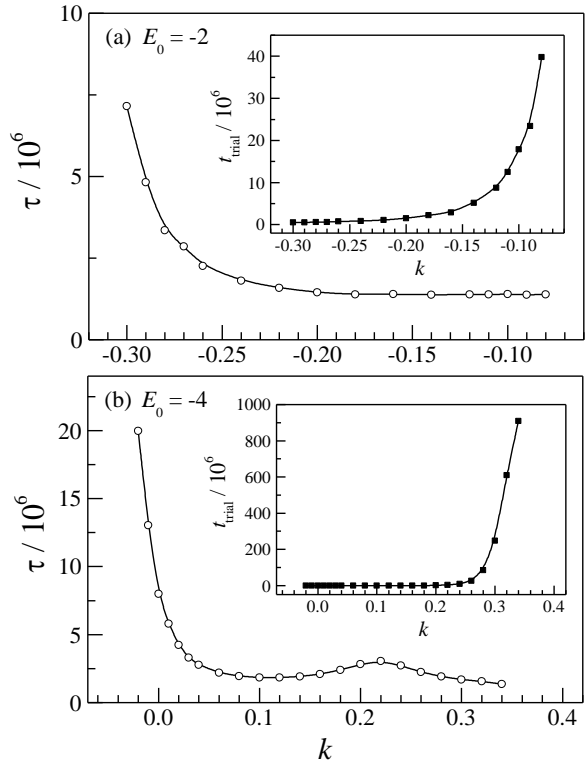

**Fig. 7** Plot of the translocation time τ versus the energy gradient $k$ for polymer length at $E_0 = -2$ **(a)** and $E_0 = -4$ **(b)**. The insets show the dependence of the trial time $t_{trial}$ on $k$. Simulation parameters: polymer length $N = 64$, pore length $L = 10$, driving force $f = 0.1$.

In short, we find that the calculation based on the Fokker-Planck equation and the results of the Monte Carlo simulations produce qualitatively similar results on the dynamical behavior of the translocation time. The disagreement between the two methods might be due to the quasi-equilibrium assumption of the Fokker-Planck equation but an out-of-equilibrium process in the simulation. The out-of-equilibrium process is an important characteristic of polymer translocation [28]. The reason for the out-of-equilibrium of translocation is attributed to the fact that the translocation time of the forced translocation is shorter than the configuration relaxation time of the polymer, which was observed clearly in many simulations [28, 50-52]. The out-of-equilibrium process of polymer translocation is a big challenge for theoretical calculations. Moreover, it was pointed out that the diffusion coefficient $D$ might not be a constant, but rather dependent on the position of the polymer inside the nanopore [53]. The exact dependence is not known because of the complexity of the system. If $D$ is only a function of position, then the behavior of translocation time is still only dependent on the free-energy landscape. Therefore, the position-dependent diffusion coefficient could change the translocation time but not the behavior of the translocation time. A further consideration in the theoretical calculation is that $D$ might be dependent on the interaction between the polymer and the pore. Simulation observed that $D$ decreases with increasing attraction of the surface for polymer chains inside a narrow slit [54] as well as inside a long tube [55]. Therefore, computer simulation is still an important tool to understand the translocation process of a polymer through a nanopore. In the simulations, the position-dependent and interaction-dependent diffusion properties are naturally included. Our results imply that the position-dependent diffusion coefficient does not change the behaviors of polymer translocation.

Experiments found the translocation time of a polymer could be tuned by changing the polymer-pore interaction [45] or the polymer-pore interaction gradient [16]. Our theoretical and simulation results are generally consistent with the experimental results. However, we would like to point out that the three different behaviors of the translocation time discovered in this work were not observed in experiments. As there are several methods to control the polymer-pore interaction, for example, pH difference between the *cis* and *trans* sides [16], surface modification [44, 45, 56], and surface biased voltage [46, 57], we believe our theoretical system will be realized and these theoretical findings will be observed in future experiments.

## 4. Conclusions

The translocation time of a polymer chain through an energy gradient nanopore was calculated by the Fokker-Planck equation with double-adsorbed periodic boundary conditions and simulated by the Monte Carlo method. The two different methods produced qualitatively similar results on the dynamical behaviors of the translocation time. We observed three different behaviors for the translocation time, depending on the polymer-pore interaction and driving force. The results can be explained qualitatively from the free-energy landscapes of the polymer translocation. Results show that the energy gradient nanopore could accelerate or decelerate the translocation of polymer chains; such a mechanism may play an important role in the control of DNA motion through nanopores.

**Acknowledgements** This work was supported by the National Natural Science Foundation of China (Grant Nos. 11374255 and 11674277) and the Zhejiang Provincial Natural Science Foundation of China (Grant No. LQ14A040004).